\documentclass[aps,prl,twocolumn,showpacs,preprintnumbers,
amsmath,amssymb]{revtex4}
\usepackage{graphicx}
\usepackage{bm}

\begin{document}

\title{Observations of two-fold shell filling and Kondo effect in a graphene
nano-ribbon quantum dot device}

\author{C. L. Tan, Z. B. Tan, K. Wang, L. Ma, F. Yang, F. M. Qu,
J. Chen}
\author{C. L. Yang}
\email[Corresponding authors.\\]{ycl@iphy.ac.cn; lilu@iphy.ac.cn.}
\author{L. Lu}
\email[Corresponding authors.\\]{ycl@iphy.ac.cn; lilu@iphy.ac.cn.}

\affiliation{
Daniel Chee Tsui Laboratory, Beijing National Laboratory for Condensed Matter Physics, 
Institute of Physics, Chinese Academy of Sciences, Beijing 100190, People's Republic of China}

\date{\today}

\begin{abstract}
A graphene nanoribbon (GNR) with orientation along its principle axis
was obtained through a mechanical tearing process, and a quantum dot
device was fabricated from the GNR. We have studied the transport
property of the GNR quantum dot device down to dilution refrigerator
temperatures. Two-fold charging periodicity was observed in the
Coulomb-blockade measurement, signaling a shell-filling process with
broken valley degeneracy. In one of the smaller Coulomb diamonds,
Kondo-like resonance were observed, with two conductance peaks
displaced symmetrically from the zero bias voltage. The splitting of
Kondo resonance at zero magnetic field suggests spin-polarization of
the quantum dot, possibly due to the edge states of a zigzag GNR.

\end{abstract}



\maketitle

The discovery of quantum Hall effect in graphene
\cite{Novoselov_2005,Zhang_2005} has triggered a lot of studies on
this material. Attention has also been paid to the nanoribbon forms
of the material, due to their intriguing physical properties for
basic research and potential applications in future nano-electronic
circuits \cite{Nakada, Wakabayashi, Brey, Peres, Lau, Son, Pisani,
Xiao, Liu, Han, Fuhrer, Rycerz, Li}. As an unrolled version of
single-walled carbon nanotubes, GNRs could be flexibly tailored to
various shapes with controlled width, length, orientation and even
intra-molecular junctions that are suitable as the building blocks
for electronic circuits. With the unrolled structures, however, the
edges of GNRs are expected to play an important role, introducing
new physics different from carbon nanotubes. Theoretical studies
\cite{Nakada, Wakabayashi, Brey, Peres, Son, Pisani} show that GNRs
with zigzag edges have spin polarized edge states, and are
half-metals under a large transverse electric field. Similar
prediction has also been made for bilayer zigzag GNRs
\cite{Castro_2008,Sahu}. Experimentally, GNRs have been
fabricated through a reactive ion etching method with the help of
electron beam lithography \cite{Han}, or via a chemical method
\cite{Li}.  Electron transport measurements reveal that these GNRs 
are all semiconducting, with a gap inversely proportional to their width\cite{Han, Li}. 
In the experiments of Ref. \cite{Li}, sub-10-nm wide GNRs along principle axes were 
also reported. However, there is no systematic experimental study
on GNRs with identified edge orientations.

In this paper, we report our electron transport measurement on a quantum dot
(QD) device made of a very-long GNR along one of the principle axes
of a graphene sheet. Coulomb blockade diamonds with two-fold
charging periodicity were observed. Moreover, inside one of the
small diamonds a fine peak-dip-peak conductance structure was
discerned around zero bias voltage. We believe that these results
indicate the breakdown of both the valley degeneracy and the spin
degeneracy at different energy scales in the GNR.

We use the ``Scotch Tape" technique \cite{Novoselov_PNAS} to
exfoliate highly ordered pyrolytic graphite (HOPG) onto
degenerate-doped \emph{p}-type Si substrates with a 100 nm thick
SiO$_2$ layer. Besides obtaining large pieces (several $\mu$m$^2$
and up) of graphene \cite{mobilitynote}, occasionally we can get
very long GNRs. Figure 1(a) is a scanning electron microscope image
of the GNR used in this experiment. It is about 12 microns long and
seems to have a uniform width over its entire length. Atomic force
microscope (AFM) shows that the GNR has a width of $w=60$ nm and a
height of $\sim$ 1.3 nm. The height of the GNR indicates that
it has at most two graphite layers \cite{Li}.

The fact that some of the mechanically exfoliated GNRs have a
constant width over a very long distance would suggest that they
have two extremely parallel and smooth edges. This is most possible
when both edges are along one of the principle axes of the graphene
sheet.  The existence of favorable tearing directions was reported
by Geim's group on large pieces of graphene
\cite{THE_RISE_OF_GRAPHENE}, and by Dai's group on some of their
chemically derived GNRs\cite{Li}. As shown in Fig. 1(a), besides the
very long GNR, there is another GNR whose direction changes exactly
by 30 degrees across a ``kink'' structure, presumably from zigzag to
armchair or vice versa. The ``kink'' provides further evidence that
mechanically exfoliated GNRs are likely along the principle axes.

\begin{figure}
\includegraphics[width=0.62\linewidth]{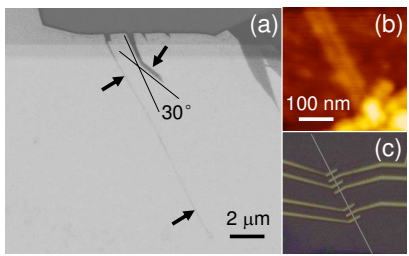}
\caption{\label{fig:Tanfig1} {(color online) (a) Scanning electron
microscope image of mechanically-torn graphene nanoribbons (GNRs).
The short GNR kinks exactly by 30 degrees. (b) Atomic force
microscope image of the long GNR near an electrode, indicating the
GNR width of about 60 nm. (c) Optical image of the electrodes
fabricated onto the long GNR (a line is drawn to represent the
GNR).}}
\end{figure}

Pd electrodes of 50-nm thick, 500-nm wide and 500-nm apart were
patterned onto the GNR using standard electron beam lithography
techniques (Fig. 1(c)). Two-probe differential conductance between
neighboring electrodes of the device was measured using
low-frequency lock-in techniques with an ac modulation voltage of 10
$\mu$V. The measurements were performed in a Quantum Design PPMS
system with temperatures $T$ down to 2 K, and in a dilution
refrigerator with $T$ down to 20 mK.

Figure 2(a) shows the differential conductance $G=dI/dV_{\rm bias}$
in a color scale as a function of bias voltage $V_{\rm bias}$ and
back-gate voltage $V_{\rm g}$, measured from one of the segments of
the GNR at $T=2$ K in the PPMS. The results measured from other
segments are qualitatively the same, and are similar to those
reported previously \cite{Han,Stampfer, Stampfer2}. The conductance is
suppressed at small $V_{\rm bias}$ due to the Coulomb blockade (CB)
effect, and the suppression is enhanced in the vicinity of the Dirac
point centered at $V_{\rm g}\sim$3 V, where the carrier
concentration is the lowest. In Fig. 2(b) we show the zero bias
CB oscillations near the Dirac point in the range $V_{\rm g}=3.24 -
3.46$ V, measured in a dilution refrigerator at $T=600$ mK and 20
mK, respectively. The conductance peaks become much sharper at 20
mK. Figure 2(c) shows the stability diagram at this temperature, in
the same $V_{\rm g}$ window as in Fig. 2(b). The most salient
feature is the two-fold charging periodicity with varying $V_{\rm
g}$. The bigger CB diamonds are about 23 meV wide, and the smaller
ones are about 15 meV wide. The height of the smaller diamonds is
around 0.6 meV. Excited states at an energy scale of $\sim 0.3$ meV
can be recognized at the edges of the diamonds, as marked by the
arrows in Fig. 2(c).

\begin{figure}
\includegraphics[width=0.75\linewidth]{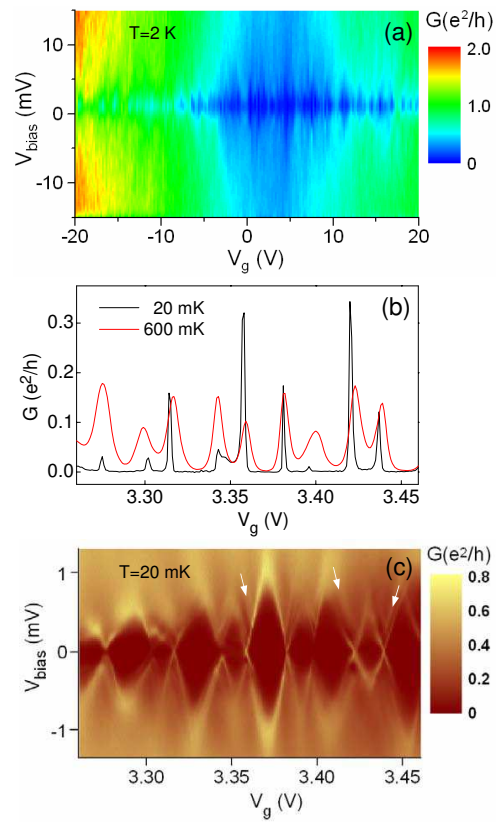}
\caption{\label{fig:Tanfig2} {(color online) (a) Color image of
differential conductance, $G=dI/dV_{\rm bias}$, as a function of
bias voltage $V_{\rm bias}$ and back gate voltage $V_{\rm g}$,
measured at $T=$2 K. (b) Coulomb oscillation of $G$ at $T=600$ mK
and 20 mK. (c) Stability diagram of the GNR device at 20 mK and in a
$ V_{\rm g}$ window from 3.26 to 3.46 V. Coulomb diamonds with a
two-fold periodicity along with the excited states outside the
diamonds (indicated by the arrows) are observed.}}
\end{figure}

The appearance of CB diamonds with excited states indicates that the
GNR forms a QD at low temperatures. The effective size of the QD can
be estimated from the width of the smaller CB diamonds, $\Delta
V_{\rm g,s}=e/ C_{\rm g}\sim 15$ meV, where $e$ is the electron
charge and $C_{\rm g}$ the capacitance between the QD and the back
gate. Assuming a simple parallel plate capacitance, the length of
the GNR QD estimated is about $510$ nm that happens to be
very close to the distance between the contact electrodes $L=500$ nm.
However, a more accurate model taking into account the finite size effect of
the GNR plate\cite{Gelmont1995} gives an effective length $L^{*
}\approx 175$ nm. The difference between $L$ and $L^{*}$ may reflect
the existence of long depletion regions that serve as
tunnel barriers near the electrodes\cite{carrier_depletion_in_CNT}.
We rule out the possibility that the device contains multiple QDs, because no joint
CB diamonds are observed.

\begin{figure}
\includegraphics[width=1.0\linewidth]{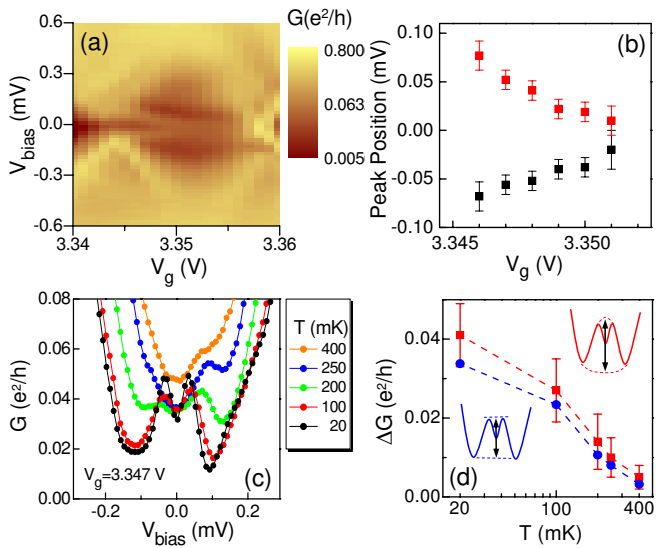}
\caption{\label{fig:Tanfig3} {(color online) (a) Color-scale plots
of the conductance $G$ in one of the Coulomb diamonds with odd
number of carrier occupation (the sixth diamond from the right in
Fig. 2(c)). Conductance peak-dip-peak structure near zero bias
voltage and its evolution with $ V_{\rm g}$ can be resolved. $T=20$
mK. (b) Peak position as a function of $ V_{\rm g}$. (c) $G-V_{\rm
bias}$ curves measured at different temperatures. The peak-dip-peak
structure disappears above $\sim$ 400 mK. (d) The amplitude of
conductance enhancement $\Delta G$ in (c) as a function of
temperature. Insets of (d) depict two different ways of estimating
$\Delta G$.}}
\end{figure}

Apparently, the two-fold periodicity observed in Fig. 2(c) resembles
the shell-filling process of a bar-like QD, where the orbital
degeneracy is lifted by the spatial confinement while the spin
degeneracy is still held. Shell-filling feature is widely observed
in QD devices made of carbon nanotubes \cite{Cobden98, Liang,
Cobden} or semiconductors\cite{Tarucha}. When the carrier occupancy
of the QD, $N$, is odd, addition of one carrier only needs to
overcome the Coulomb energy, whereas it has to pay an extra energy
to occupy the upper shell when $N$ is even. As a result, the size of
the CB diamonds shows two-fold periodicity, with the small diamonds
correspond to odd $N$ and the larger ones correspond to even $N$. As
it is known, the electrons in 2D graphene exhibit a four-fold
degeneracy near the Dirac point, i.e., two-fold spin degeneracy plus
two-fold $K$ and $K'$ valley degeneracy. The observation of a
two-fold charging periodicity possibly signals the lifting of the
valley degeneracy due to the existence of edges in the GNR
\cite{Nakada, Brey, Son, Castro_2008}.

Another interesting feature of the data is the appearance of a
Kondo-like conductance enhancement in one of the Coulomb diamond
centered at $V_{\rm g}=$3.35 V (the sixth diamond from the right in
Fig. 2(c)). Since this diamond is smaller than its neighbors, the
electron occupancy $N$ on the QD is presumably odd, leaving an
uncompensated electron spin on the QD. It seems that this
uncompensated electron spin mediates a resonant screening process
between the conduction electrons in the electrodes, thus provides a
conduction mechanism in the CB regime\cite{Kondo_Glazman,Kondo_Ng, Kondo_Wan,Meir},
as shown in Fig. 3.

Kondo resonance usually appears as a single conductance peak at zero
bias voltage (Fig. 4(a)). However, the conductance enhancement
observed here exhibits a peak-dip-peak structure around zero bias.
The two peaks are $\sim 0.1$ meV apart at the most left side of the
diamond, getting closer with increasing $V_{\rm g}$, and apparently 
merging together in the right half of the diamond\cite{note_kondo_merging}.

Usually the height of a Kondo conductance peak follows a logarithmic
temperature dependence below the Kondo temperature, and saturates
gradually at very low temperatures. The temperature dependence data 
of the peak-dip-peak structure are shown In Fig. 3 (c). 
As can be seen in Fig. 3(d), the amplitude of the overall structure, $\Delta G$, 
extracted using two different ways (depicted in the insets), does roughly
follow a logarithmic temperature dependence below 400 mK.

It is known that the Kondo resonance peak splits into two peaks that
are symmetrically displaced from the zero bias if there is a
preferred spin orientation (i.e., when the spin degeneracy is
lifted, as illustrated in Fig. 4(b)) \cite{Meir}. Such splitting has
been observed in both 2D semiconductor QDs and carbon nanotubes,
where the spin degeneracy is lifted by either an applied magnetic
field \cite{Goldhaber} or by the exchange coupling to ferromagnetic
contacts \cite{Hauptmann}. The lifting of spin degeneracy in this
experiment is surprising in that there is neither external magnetic
field nor ferromagnetic contacts.

Although the exact nature is obscure, the observed peak-dip-peak
structure may be qualitatively understood by assuming some degree of
spin polarization on the GNR QD itself. As aforementioned, preferred
spin orientation is expected at the edges of a zigzag GNR \cite{Son,
Pisani, Castro_2008, Sahu}. In following paragraphs, by analyzing the
CB data in Fig. 2 we will show that the GNR measured in this experiment is
quite likely to be a zigzag one.

Let us first show that the QD is unlikely made of an armchair GNR.
The band dispersion of an armchair GNR takes the same Fermi velocity
($v_F=1.0\times 10^6$ m/s) as in a bulk graphene, except a gap near
the neutral point in the semiconducting cases. Taking $L^{*}\approx
175$ nm, the level spacing due to longitudinal confinement for this
bar-like GNR is $\epsilon_L=hv_F/2L^{*}\approx 12$ meV, and the
energy spacing between subbands is $\epsilon_w=hv_F/2w\approx 34$
meV. In order to explain the excited state energy of $\Delta\approx
0.3$ meV as observed in the CB measurement, one has to assume that
there are many subbands across the Fermi level. Simple
estimation\cite{excited-energy} shows that the Fermi energy
$\epsilon_F\sim 430$ meV, and the number of carriers in the GNR
$N\sim710$. This would lead to a Dirac point (the neutral point)
that is located 20 V away from the $V_{\rm g}$ window of Fig. 2 (c).
However, the experimental data indicate that the Dirac point is
located at several volts of $V_{\rm g}$, or at worst within 10 V
from the $V_{\rm g}$ window. In addition, if the GNR is armchair, in
the regime near the Dirac point there should be at least a few big
CB diamonds whose height is comparable to the confinement energy
spacing of 12 meV and whose width is wider than 250 meV in $V_g$.
However, as shown in Fig. 2 (a), there is no CB diamond having these
large sizes.

On the other hand, for a zigzag GNR, the energy bands near the Fermi
level is nearly flat due to the existence of edge states. Therefore,
the confinement along its length direction yields a relatively small
energy level spacing, so that the lack of large CB diamonds is
naturally understood. To estimate the level spacing (the excited
states energy), we recall that the bands of edge states are slightly
dispersive because of the next nearest-neighbor interaction
\cite{Neto_2006}, with a small group velocity $v'_F$ at the order
$v'_F/v_F\sim t'/t\sim 0.03$, where $ t\approx 3.0$ eV and $t'\sim
0.1$ eV are the first and the second nearest-neighbor hopping
energy\cite{Reich_2002}, respectively. The level spacing estimated
from the dispersion of edge states is roughly $\epsilon'_L\approx
hv'_F/2L^{*}\sim 0.4$ meV, consistent with the observed values.
Therefore, we believe that our GNR is most likely to be a zigzag GNR rather than an armchair
one.

\begin{figure}
\includegraphics[width=0.9\linewidth]{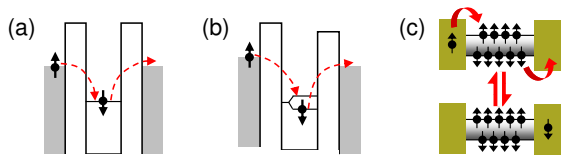}
\caption{\label{fig:Tanfig4} {(color online) (a) Elastic Kondo
resonance in a quantum dot device provides a conduction mechanism at
zero bias voltage in the Coulomb blockade regime. (b) Inelastic
Kondo resonance due to lifted spin degeneracy leads to a pair of conductance peaks
at finite bias voltages (symmetrical about zero bias).
(c) A toy model depicts the possible Kondo resonance in a zigzag GNR.
The spins are parallel on one edge but are antiparallel between opposite edges.
Kondo resonance becomes inelastic when the electrostatic environment of the two edges is
different.}}
\end{figure}

Based on the spin configuration of the edge states in a zigzag GNR
\cite{Son, Pisani} that the spins on the same edges are parallel but
on opposite edges are antiparallel, we present a toy-model in Fig. 4
(c) to account for the lifting of spin degeneracy. In this model,
electrons with different spin will fill onto different edges. When
the local electrostatic environment of the two edges are different,
filling to different edge will cost different energy. This would
lead to the observed spin splitting in odd $N$ GNR QD. In this
scenario, the gate-voltage dependence of the Kondo peak splitting
can also be naturally explained through the dependence of local
electrostatic environment on the gate voltage. In fact, it is almost
unavoidable to have randomly trapped charges in SiO$_2$ surrounding
the GNR. The distribution of these charges, thus their stray field,
is tunable by applying a gate voltage. To further clarify the issue,
more experiments such as magnetic field dependence of the peak
splitting are demanded.

To summarize, we have observed two-fold charging periodicity in a QD
device made of a GNR. We also observed Kondo-like conductance peaks
in one of the Coulomb diamonds. The observation of Kondo splitting
at zero magnetic field suggests the lifting of spin degeneracy
due to spin polarization on the QD itself, in favor of theoretical
predictions on a zigzag GNR where the spin polarization is
due to its highly degenerate edge states.

We would like to thank H. F, Yang and X. N. Jing for experimental
assistance, Y. Q. Li, Q. F. Sun, W. J. Liang, and K. Chang for
helpful discussions. This work was supported by the NSFC, the
National Basic Research Program of China from the MOST, and by the
Knowledge Innovation Project of CAS.


\begin{thebibliography}{10}
\bibitem{note_kondo_merging} It is uncertain whether the resonance 
exhibits a true single peak in this region, due to insufficient resolution of the current data.
\bibitem{excited-energy} In the CB measurements where the carrier occupancy
$N$ is large, the energy of excited states satisfies
$\Delta=1/g(\epsilon_F)$, with $g(\epsilon_F)$ the density of states
at the Fermi level. When many subbands are occupied, $g(\epsilon_F)$
can be estimated the same way as for a 2D graphene:
$g(\epsilon_F)=\sqrt{2\pi NS}/(\pi\hbar v_{F})$, where $S$ is the
area of the QD.
\end{thebibliography}
\end{document}